\documentclass[conference]{IEEEtran}

\usepackage{cite}
\usepackage{amsmath,amssymb,amsfonts}
\usepackage{algorithmic}
\usepackage{graphicx}
\usepackage{textcomp}
\usepackage[table]{xcolor}
\usepackage[center]{caption}
\usepackage{pifont}
\usepackage{amssymb}
\usepackage{multirow}
\usepackage{dblfloatfix}
\usepackage{graphicx}
\usepackage[linesnumbered,ruled]{algorithm2e}
\usepackage{subcaption}
\usepackage{tabu}
\usepackage{makecell}
\usepackage[left=0.64in,right=0.64in,top=0.74in]{geometry}

\usepackage{soul}

\begin{document}

\title{Combining Software Defined Networks and Machine Learning to enable Self Organizing WLANs\\}

\author{\IEEEauthorblockN{1\textsuperscript{st} \'Alvaro L\'opez-Ravent\'os}
\IEEEauthorblockA{\textit{Information and Communication Technologies} \\
\textit{Universitat Pompeu Fabra (UPF)}\\
Barcelona, Spain \\
alvaro.lopez@upf.edu}
\\
\IEEEauthorblockN{3\textsuperscript{rd} Sergio Barrachina-Mu\~{n}oz}
\IEEEauthorblockA{\textit{Information and Communication Technologies} \\
\textit{Universitat Pompeu Fabra (UPF)}\\
Barcelona, Spain \\
sergio.barrachina@upf.edu}
\and
\IEEEauthorblockN{2\textsuperscript{nd} Francesc Wilhelmi}
\IEEEauthorblockA{\textit{Information and Communication Technologies} \\
\textit{Universitat Pompeu Fabra (UPF)}\\
Barcelona, Spain \\
francisco.wilhelmi@upf.edu}
\\
\IEEEauthorblockN{4\textsuperscript{th} Boris Bellalta}
\IEEEauthorblockA{\textit{Information and Communication Technologies} \\
\textit{Universitat Pompeu Fabra (UPF)}\\
Barcelona, Spain \\
boris.bellalta@upf.edu}
}

\maketitle

\begin{abstract} 
Next generation of wireless local area networks (WLANs) will operate in dense, chaotic and highly dynamic scenarios that in a significant number of cases may result in a low user experience due to uncontrolled high interference levels. Flexible network architectures, such as the software-defined networking (SDN) paradigm, will provide WLANs with new capabilities to deal with users' demands, while achieving greater levels of efficiency and flexibility in those complex scenarios. On top of SDN, the use of machine learning (ML) techniques may improve network resource usage and management by identifying feasible configurations through learning. ML techniques can drive WLANs to reach optimal working points by means of parameter adjustment, in order to cope with different network requirements and policies, as well as with the dynamic conditions. In this paper we overview the work done in SDN for WLANs, as well as the pioneering works considering ML for WLAN optimization. Finally, in order to demonstrate the potential of ML techniques in combination with SDN to improve the network operation, we evaluate different use cases for intelligent-based spatial reuse and dynamic channel bonding operation in WLANs using Multi-Armed Bandits.
\end{abstract}

\begin{IEEEkeywords}
SDN, Machine Learning, Wireless, Spatial Reuse, Channel Allocation.
\end{IEEEkeywords}


\section{Introduction}\label{sec:intro}

In recent years, IEEE 802.11-based WLANs, commonly known as Wi-Fi networks, have experienced a remarkable growth in terms of traffic consumption. According to \cite{index2015cisco}, in 2016 more traffic was offloaded from cellular networks onto Wi-Fi than remained on them. Moreover, they expect by 2021 that the 63 \% of total mobile data traffic will be offloaded onto Wi-Fi network as a consequence of an increased use of portable and handheld devices. In this context, network capacity needs to be targeted to cope with the expected data traffic. Thus, efforts are focused in network densification as the spectrum scarcity and the high spectral efficiency achieved by current technologies are limiting factors \cite{bellalta2016ieee}.

Regarding dense deployments, there exist some potential issues in regards of performance degradation. Existing channel access protocols, such as carrier sense multiple access (CSMA), have been designed to operate efficiently in non-dense scenarios, and they may become a bottleneck when pushed further. In dense WLANs, due to the great number of contending nodes using CSMA, we could find three well-known performance issues. We refer to the hidden and exposed node problems, and to the flow starvation. In terms of performance, the appearance of these issues can cause a remarkable degradation of the experienced throughput due to different factors, such as a large number of collisions or wasting useful time slots. Moreover, some solutions like request-to-send and clear-to-send (RTS/CTS) mechanisms that are intended to avoid the hidden and exposed node problems, can lead to an excessive control packet overhead, which may negatively affect the overall performance, too. Apart from the above-mentioned, other concerns are related with chaotic deployments since they lead to have an excessive co-channel and adjacent channel interference (CCI/ACI) levels, directly caused by the lack of frequency planning and inefficient power configuration choices.

To cope with the aforementioned challenges, the software-defined networking (SDN) paradigm can be applied to Wi-Fi networks in order to enable a more efficient and flexible network control and management. The main concept behind SDN is that it proposes to decouple the control and the data planes into different layers, with a central controller performing configuration changes with a global view of the network state. As a result, control processes are removed from forwarding devices, which stand as simple programmable nodes that directly depend on the controller's instructions. In consequence, networks can be adjusted dynamically according to the knowledge extracted from statistics, which are collected at the central entity. This specific characteristic of SDN is quite relevant for wireless environments due to their non-stationary conditions (i.e., users moving, diverse traffic requirements and changing channel conditions). Having a dynamic and centralized control design, the overall performance of the network can be improved and  interferences, unbalanced situations or system failures mitigated. In this regard, network management and data analytics play a key role in order to increase network efficiency. For instance, network information such as the signal-to-interference-plus-noise ratio (SINR), the received signal strength indicator (RSSI), the total number of active users and throughput rates can be easily collected. Thus, network optimization needs to exploit this useful data and the use of learning algorithms can lead this processes. This envision open up new research directions and so, we focus our studies in the joint integration of machine learning and SDN for wireless optimizations.

The rest of the paper is organised as follows: in Section \ref{sec:sdn_sdwn}, we present a general overview about the SDN paradigm, reviewing current implementations of the SDN architecture into wireless networks. Then, we point out different features to be taken as future research directions. After, in Section \ref{sec:intelligent}, we discuss an architecture involving wireless SDN and ML solutions, together with an overview of different management functionalities. Later, in Section \ref{sec:use_case}, we perform a proof of concept with the aim to demonstrate how ML can enable self-organizing WLANs. Through different use cases, we evaluate the usage of ML over SDN-controlled WLANs with the aim to to find the best configuration according to a Max-min fairness policy. To do so, we exploit the Multi-Armed Bandits (MABs) framework to empower a collaborative behavior between them. Finally, conclusions are stated in Section \ref{sec:conclusions}.


\begin{table*}[b]
\centering
\caption{Taxonomy of the related work presented}
    \begin{tabular}{|c|c|c|c|c|c|c|}
        \hline
         \makecell{\textbf{Name}} & \makecell{\textbf{Application} \textbf{development}} & \makecell{\textbf{Southbound} \textbf{communication}} & \makecell{\textbf{VAP / LVAP}} & \makecell{\textbf{Separated MAC}} & \makecell{\textbf{End-user} \textbf{modification}}\\ 
         \hline
         OpenRoads & \ding{51} & OpenFlow + SNMP & \ding{51} & \ding{55} & \ding{55}\\
         \hline
         Odin & \ding{51} & OpenFlow + Proprietary & \ding{51} & \ding{51} & \ding{55}\\
         \hline
         CloudMAC & \ding{51} & OpenFlow & \ding{51} & \ding{51} & \ding{55}\\
         \hline
         \AE therflow & \ding{51} & Extended OpenFlow & \ding{55} & \ding{55} & \ding{55}\\  
         \hline
         COAP & \ding{51} & Extended OpenFlow & \ding{55} & \ding{55} & \ding{55}\\
         \hline
         Ethanol & \ding{51} & OpenFlow + Proprietary & \ding{51} & \ding{51} & \ding{55}\\
         \hline
         Aeroflux & \ding{51} & OpenFlow + Proprietary & \ding{51} & \ding{51} & \ding{55}\\
         \hline
         OpenSDWN & \ding{51} & OpenFlow + REST & \ding{51} & \ding{55} & \ding{55}\\
         \hline
         BeHop & \ding{51} & OpenFlow + Proprietary & \ding{51} & \ding{55} & \ding{55}\\
         \hline
         EmPOWER & \ding{51} & OpenFlow + Proprietary & \ding{51} & \ding{55} & \ding{55}\\
         \hline
    \end{tabular}
    \label{tab:Taxonomy}
\end{table*}

\section{SDN: From wired to wireless}\label{sec:sdn_sdwn}
\subsection{SDWN through SDN}

SDN is a novel network architecture paradigm that is dynamic, manageable, cost-effective and adaptable. Moreover, SDN decouples network control and forwarding functions into different planes, allowing the underlying infrastructure to be abstracted from application and network services. In consequence, unlike distributed architectures, in which forwarding devices listen for events from their neighbors and make decisions based on a local view, the network infrastructure (i.e., switches and routers) just act as packet forwarding devices. In addition, SDN empowers programmability and network function virtualization (NFV) at the controller, allowing network administrators to have flexibility and a fine-grained control over the entire network. Thus, SDN reduces capital and operational expenditures (CapEX and OpEX, respectively), while enabling innovation. 

Typically, the SDN architecture is divided into three different layers, which can be found in literature as infrastructure, control and application layers. The first one contains the different network elements that follow the rules provided by the controller. The second one involves the controller, which is in charge of configuring the devices as well as to  the different services. The last one contains the network applications which define the different policies to be applied over the network. Communication between layers is done by means of the northbound and southbound interfaces. The former is based on APIs (e.g. REST) that are intended to application development, while the latter is based on standard protocols such as OpenFlow, Simple Network Management Protocol (SNMP) or Control And Provisioning of Wireless Access Points (CAPWAP). However, none of this protocols are intended to wireless communications and therefore, as currently defined, they cannot control layer 2 traffic over wireless networks nor report measurements of the wireless medium. To overcome with that issue, modifications by means of extending the current protocols, or even the use of proprietary ones should be adopted to enable the control of wireless devices.

Although SDN needs to be clearly reconditioned in order to be used in wireless networking, the previously described features have pushed the trend to adopt SDN for WLANs. In this context, the concept of software-defined wireless networking (SDWN) appears with a clear aim to improve the management of wireless networks and so, SDWN has become an emerging research branch of SDN. Many publications have focused on identifying the concerns and applications of SDWN, as well as suggesting different network architectures. SDWN solutions go from extending the OF protocol with new messages, to the implementation of applications on top of OF controllers that have their own proprietary control messages. Next, in \ref{subsection:overview}, we review different architecture solutions proposed for SDWN.

\subsection{Overview of proposals for SDWN} \label{subsection:overview}

To begin with, OpenRoads \cite{yap2010openroads} was the first project focusing in SDN for WLANs. Moreover, it also introduced a testbed to control mobility between Wi-Fi and WiMax base stations. OpenRoads consists on a three layer-based architecture that is divided into physical, slicing and control layer. The physical layer is made of all the devices that are OF-enabled. The control layer is in charge of network orchestration and device configuration. Finally, the slicing layer intercepts OF protocol messages to support the slicing layer according to the network administrator policy. Thus, different network administrators can operate over the same physical network, since the slicing layer divides it into multiple logical networks. From here, other solutions such as Odin \cite{suresh2012towards} came up. The Odin's architecture is composed by an Odin master (running on the OpenFlow controller), and an Odin agent (running on the APs). The Odin master communicates to the switches and the APs by means of the OpenFlow protocol, in order to control the wired connections, whereas it uses a custom protocol to communicate to each Odin agent, with the aim of collecting different network statistics (i.e., RSSI, SINR, etc.). As a result, the network is able to manage mobility, load balancing and interference in wireless connections. In addition, time-critical operations (e.g. ACKs) are performed by the APs, and non-time-critical operations are handled by the controller. Regarding client-AP association, Odin implements the concept of logical virtual access point (LVAP), which are client-specific. So, each user receives a unique BSSID to be connected to. This implementation allows client isolation as well as performing a hand-off process without triggering any re-association mechanism, since LVAPs can be removed from one AP and transferred to another. However, the hand-off in Odin is still performed based only on the RSSI, which could lead to load imbalance situations. Similar to Odin, OpenSDWN \cite{schulz2015opensdwn} is a framework that introduces a more detailed wireless data-path transmission control, enabling user-service differentiation by identifying and classifying flow types. To do so, OpenSDWN uses per-client middle-boxes, called virtual middle-boxes (vMB), that can be migrated from one AP to another. Therefore, network functionalities are migrated to destination APs as the user performs a hand-off. From Odin, OpenSDWN inherits Odin's LVAPs concept as well as the mobility method and user isolation. Later on, BeHop \cite{yiakoumis2015behop} and Ethanol \cite{moura2015ethanol} appeared as other solutions in the SDWN context, which took the same basis as Odin. First, BeHop architecture consists of a central controller, a set of APs forming the data plane, and a network monitor and data collector. Each BeHop AP acts as an OpenFlow switch that contains per-client virtual APs (VAPs), and a client table to track the user information (e.g., client-VAP mapping) and the network status information (e.g., channel and power allocation). Here, the network control is performed through a BeHop own proprietary API used for channel and power allocation purposes. Moreover, through a dedicated interface, the controller is able to access the data stored in the network monitor, in order to take advantage of it and enhance network management. Regarding Ethanol, it consists of two types of devices, the controller and the Ethanol-based APs, or Ethanol agents. Ethanol uses its own proprietary code to gather link information from the APs (e.g., SINR or bit rate) in order to provide the controller with statistics for network managing. Open research directions in Ethanol aim to guarantee security and quality of service (QoS) through traffic shaping. At last, EmPOWER \cite{riggio2015programming} is an SDWN programming architecture that provides a set of Python based APIs, which model the fundamentals of wireless management. The aim of this architecture is to reduce complexity by applying four abstractions, each of one addressing a different control aspect such as: the state management, resource allocation, network monitoring, and network reconfiguration. Communication between wireless terminals and the controller is done by a proprietary protocol, whereas OpenFlow is used for managing the switching operations. Regarding time-critical actions, CloudMAC \cite{vestin2013cloudmac} proposed to break down the MAC operations by offloading them into different devices. Therefore, physical APs are in charge of time-critical MAC operations, whereas virtual APs (VAPs) are in charge of MAC generation. Besides, communication between them is performed through a layer 2 tunneling. The rest of the architecture is composed by an OpenFlow switch, which is used to forward packets between APs and VAPs, and an OpenFlow controller that orchestrates the network according to the user-defined policies. Similarly, Aeroflux \cite{schulz2014aeroflux} also promotes a separation between MAC features by implementing a 2-tier control plane. Here, the global control plane (GC) handles non-real time tasks such as authentication and load balancing, whereas the near-sighted control plane (NSC) is located closer to the APs to manage time-critical operations such as rate control and power adjustment. Then, this architecture emphasizes that control plane delays need to be short. 

\begin{figure*}[b]
\centering
	\includegraphics[width=.87\textwidth]{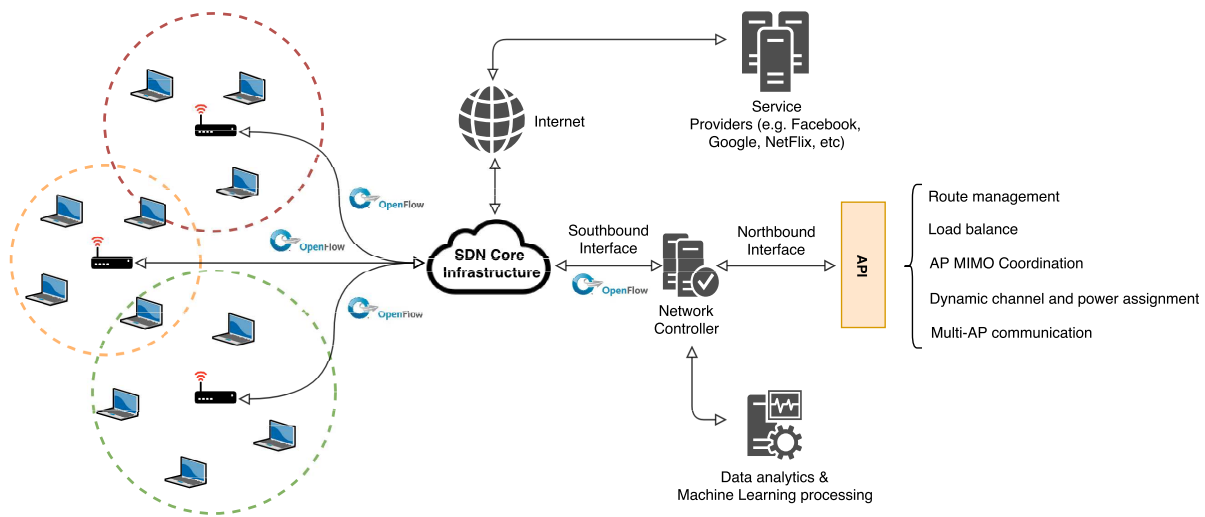}
	\caption{SDWN architecture with knowledge plane}
	\label{fig:sdn_kp} 
\end{figure*}

In contrast to the previous works reviewed, \AE therflow  \cite{yan2015aetherflow} and COAP (Coordination framework for Open APs) \cite{patro2015coap} extended the OpenFlow protocol in order to manage the communication between the controller and the APs. In consequence, both techniques simplify the data plane programmability as there is no need of extra software implementation. Thus, the extended OpenFlow protocol by itself comprises all the required messages to allow the controller gather different network statistics such as RSSI, SINR, bitrate or airtime usage.

\subsection{SDWN applications for wireless networking}

In the previous section we presented a set of different proposals. Most of them only propose or implement mechanisms to enhance mobility. However, here we present other functionalities that can be implemented:

\begin{itemize}
	\item \textbf{Spatial reuse:} power control mechanisms are essential in order to reduce interference. In SDWN environments, thanks to the centralised control plane, power control mechanisms can be applied to avoid unnecessary overlaps between WLANs. In addition, the set-up of different clear channel assessment (CCA) levels could enhance the spatial reuse.
	\item \textbf{Dynamic channel allocation (DCA):} by gathering channel statistics in the controller, SDWN can perform dynamic channel allocation to minimize co-channel interference between WLANs.
	\item \textbf{Dynamic channel bonding (DCB):} the use of channel bonding based on the spectrum occupancy of neighboring WLANs can be performed as a solution to increase throughput rates and reduce interference between nodes, allocating different channel widths to each WLAN based on its traffic demands and capabilities.
	\item \textbf{Multi-AP communication:} by taking advantage of network programmability, multiple connections per user to different APs could be easily managed. The controller would be in charge of deciding whether or not the use of multiple simultaneous connections improve user and network performance, as well as to take actions by installing new forwarding rules in the forwarding devices.
	\item \textbf{Multiple input multiple output (MIMO) and multi-user (MU) MIMO coordination:} This application is more related with a joint SDN and SDR framework. However, the programmability of SDN creates a great opportunity for SDR to be applied and therefore, techniques such as interference coordination and alignment can be implemented in order to reduce and mitigate interfering signals. Coordination of such techniques can lead future WLANs to a new level of complexity, but with high performance gains.
	%
\end{itemize}


\section{Towards intelligent networking} \label{sec:intelligent}

The SDWN paradigm is extremely flexible as networks can be dynamically reconfigured to handle new states. Thus, the introduction of machine learning techniques constitute a potential solution to achieve higher gains in terms of network performance. By using different techniques, patterns can be extracted from data sets, or learned through interacting with the environment. Therefore, the knowledge extracted from past observations can be applied to update the behavior of the network. Existing machine learning algorithms are generally classified into three different categories depending on how the learning process is done. Supervised learning (SL) algorithms are trained using labeled examples. By comparing the predicted output with the labeled ones, these algorithms update the model accordingly to the error measured. On the other hand, unsupervised learning (USL) algorithms are used against data that has no historical labels. Thus, USL algorithms try to focus on arranging samples into different groups. Last, reinforcement learning (RL) algorithms, which through trial and error, try to find the actions that yield the greatest rewards.

The inclusion of machine learning into networking motivated the consideration of a new architectural division due to the fact that this kind of algorithms does not belong to data nor control planes. The new architectural division is the knowledge plane (KP), which was proposed in \cite{clark2003knowledge}, and which intends to place machine learning techniques over the network architecture scheme. The KP is responsible for learning the behavior of the network, and the decision-making process. Basically, the KP processes the statistics collected by the control plane, transforms them into knowledge via machine learning algorithms, and uses that knowledge to make decisions. Hence, in the context of SDN networks, the KP participates actively in the network orchestration due to its interaction with the controller, which configures the network according to KP's instructions. In the literature, the joint consideration of SDN and machine learning techniques can be found as Knowledge-Defined Networking (KDN) \cite{mestres2017knowledge}. This new paradigm consists in combining data, control and knowledge planes to provide automated network control. Figure \ref{fig:sdn_kp} depicts an architecture that merges both KP and SDWN concepts to have a flexible wireless environment. Here, the SDN paradigm is identified in how the network is orchestrated since the control plane is managed by a controller that communicates and requests different information from the APs through OpenFlow. In regards to the machine learning related functionalities, a dedicated server, in which data is stored and machine learning algorithms executed, it is connected to the controller to take full advantage of network statistics to take decisions. Through the results from the machine learning algorithms, the decision-making process according to the knowledge obtained can be driven directly by the KP in an autonomous way based on a set of predefined requirements. On top of the controller, network applications are executed in order to give the directives to the controller for managing the network. In this context, some applications already done are:

\begin{itemize}
    \item \textbf{Traffic prediction and classification:} both features were the earliest machine learning applications in the networking field. In this context, traffic classification is done in order to ensure QoS as well as quality of experience (QoE). Thus, statistics gathered by the controller can be used to classify data flows into different QoS-categories. On the other hand, traffic prediction is used to forecast the total amount of traffic expected. As an example, in \cite{azzouni2017neutm}, neural networks (NN) are used to perform traffic prediction by using flow level statistics together with a learning window of past time intervals, which repetitively trains the algorithm in order to characterize and predict the network behavior. Traffic prediction solutions may lead to have proactive systems in which different actions can be triggered before traffic imbalances happen. For instance, some actions could lead to a  reconfiguration of the spectrum allocation in order to provide more bandwidth to a group of WLANs, or trigger load balance mechanisms.
    
    \item \textbf{Routing:} Regarding to the management of the wired part of the network, routing strategies have been tackled such as in \cite{kim2016congestion}, in which is proposed a network congestion prevention mechanism based on the Q-learning algorithm. In case of detecting congestion between a link pair, the algorithm recomputes the reward matrix accordingly to the inputs, in order to search a new route. As the authors proved, in comparison with Dijkstra's algorithm, Q-learning based routing provides better results.
    
    \item \textbf{Security:} This is one of the most important factors that SDN architectures must face. The centralized nature of the control plane has many benefits, but it is a risky approach in terms of security, as all the network control is placed in a single point. For instance, current attacks such as denial of service (DoS) can be potentially critical, since the control plane is no longer distributed, and so the entire network can be compromised. In this context, machine learning can help to achieve a good level of security due to its ability to automatically find correlations in data. Deep learning techniques, such as the ANN proposed in \cite{tang2016deep}, are good mechanisms to detect any anomaly by just analyzing few per flow statistics. So, the algorithm compares any incoming traffic with the previous ones and raises an alert when the deviation between them is greater than a certain threshold. In consequence, attacks such as DoS can be detected and mitigated.
    
    \item \textbf{Spatial reuse and channel bonding:} these are two techniques that are gaining attention since last IEEE 802.11ax amendment supports both of them. The former is based on the application of different techniques such as transmission power control (TPC) and CCA adjustment in order to control the potential drawbacks of uncoordinated deployments. The later refers to a technique in which two or more adjacent channels, within a given frequency band, are temporally combined to increase throughput and data transfer between devices. The application of this techniques have opened a new set of challenges in wireless environments and so, different works attempted to enhance the network performance by their application. First, in the work done in \cite{wilhelmi2019collaborative}, MABs are used for finding the AP configuration that maximizes the aggregate throughput. There, the authors analyze different policies, in which the nodes' learning process is done by means of exploiting and exploring the medium. In regards of DCB, the work done in \cite{barrachina2019dynamic}, assesses the problem for dense WLANs by evaluating different DCB policies. There, the authors show, through analytical results, that always selecting the widest available bandwidth is counterproductive in the long term. Moreover, authors conclude that, in non-fully overlapping scenarios, the optimal solution is to apply different policies depending on the context of each WLAN, and therefore they must be on-line learned.
\end{itemize}

\begin{figure*}[b]
\centering
     \begin{subfigure}[]{.3\textwidth}
         \centering
         \includegraphics[ width=.8\textwidth]{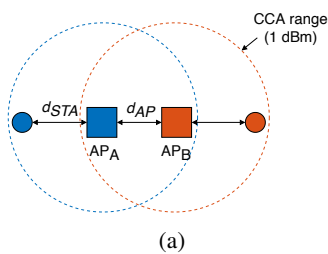}
         \caption{}
         \label{fig:scenario2}
     \end{subfigure}
	 \begin{subfigure}[]{.3\textwidth}
         \centering
         \includegraphics[ width=\textwidth]{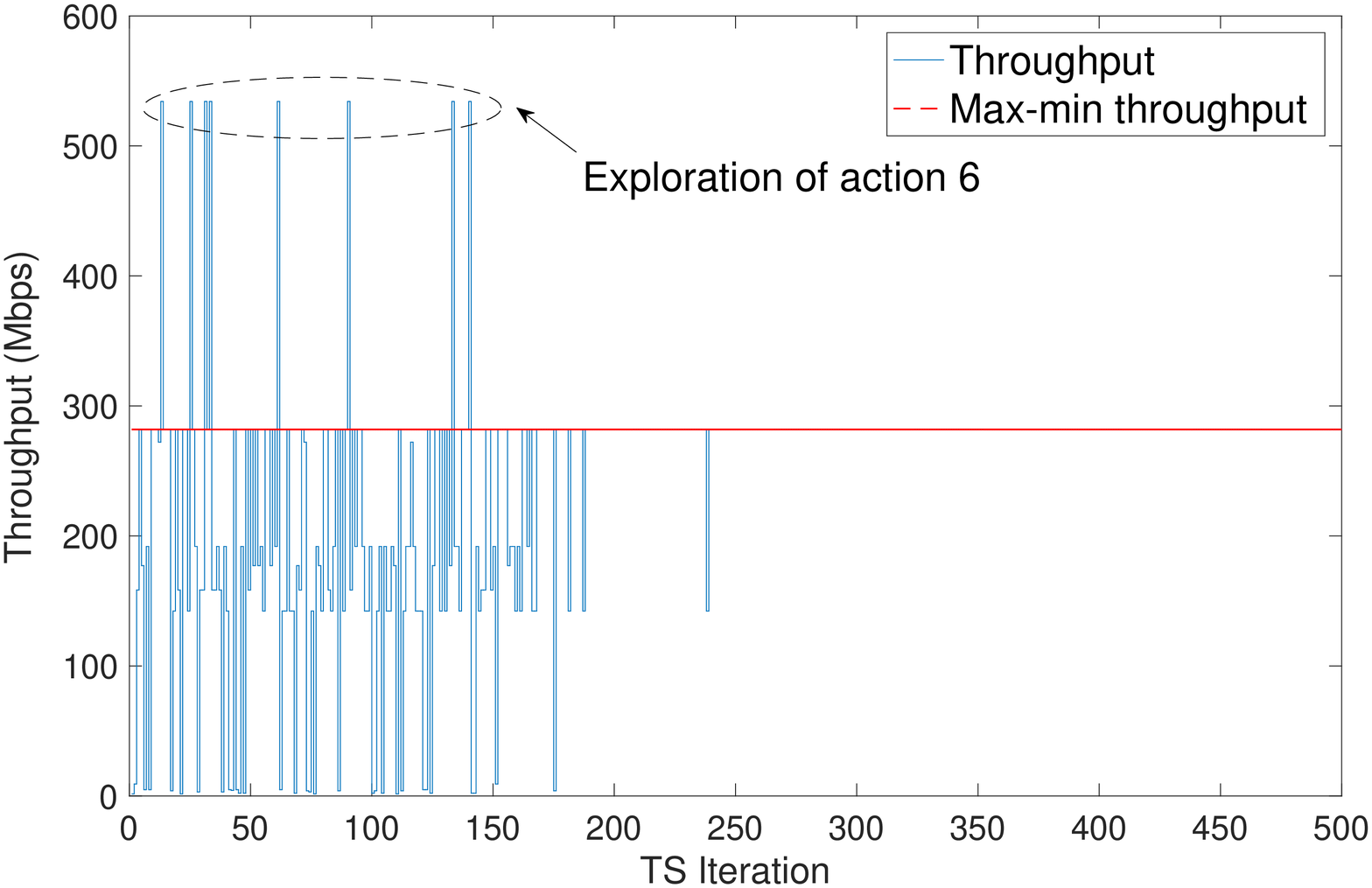}
         \caption{}
         \label{fig:Evolution2}
     \end{subfigure}
     \begin{subfigure}[]{.3\textwidth}
         \centering
         \includegraphics[ width=.95\textwidth]{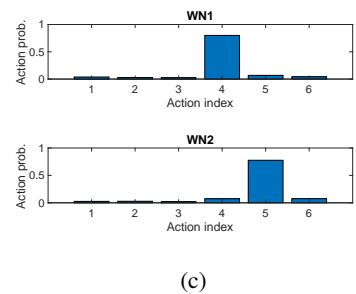}
         \caption{}
         \label{fig:actions2}
     \end{subfigure}
     \label{fig:experiment1}
     \caption[]
     {\small Use case with 2 WLAN. (a) Scenario considered. (b) Evolution of the throughput experienced by WLAN A. (c) Histogram of the probabilities for each action}
\end{figure*}

\section{Performance evaluation} \label{sec:use_case}

In order to assess the integration of the KP, we have studied the application of machine learning algorithms to tackle the spatial reuse and channel bonding issues. To do so, we have considered an SDWN composed of different WLANs, whose APs' power and channel configurations are defined by the ML server, and then advertised by the controller. By including intelligent operations, we expect to increase the network performance, aiming to find the best configuration according to a policy, while empowering a collaborative behavior\footnote{All the simulations have been performed using the SFCTMN framework developed in \cite{barrachina2019dynamic} and the learning package used in \cite{wilhelmi2019collaborative}.}. In this context, the problem is modelled through the multi-armed bandits (MABs) framework by defining a set of $K$ configurations, which correspond to any combination of channel range and transmit power that each WLAN can select (refer to Table \ref{tbl:simulation_parameters}). Moreover, as an action-selection strategy, we use Thompson sampling (TS) algorithm, since it has been shown to provide better results than other well-known algorithms such as Upper Confidence Bound (UCB) for similar problems in WLANs \cite{wilhelmi2019collaborative}. The TS algorithm is a Bayesian algorithm that constructs a probabilistic model of the rewards observed by each configuration. After selecting an arm to play, TS observes the reward, and updates its prior belief in a way that the probability of a particular arm being optimal matches with the probability of each arm being selected. In practice, this is done by sampling each arm from its posterior distribution, and selecting the one that returns the maximum expected reward. Accordingly, it randomly selects the probabilistic optimal configuration. Algorithm \ref{alg:thompsons} shows in detail the implementation of TS for this use case.

\begin{table}[t]
	\centering
	\caption{\small{Simulation parameters and action mapping}}
	\resizebox{\linewidth}{!}{%
		{\begin{tabular}{|c|c|c|}
				\hline
				\textbf{Parameter} & \textbf{Description} & \textbf{Value} \\
				\hline
				$C$ & Set of channels & 1 / 2 / 3 / 4 \\ \hline
				$P_{tx}$ & Set of transmit power values & 1 dBm / 20 dBm \\ \hline
				$f$ & Central frequency & 5 GHz\\ \hline
				$B$ & Bandwidth & 20 MHz\\ \hline
				$\text{SUSS}$ & Spatial streams per user & 1 \\ \hline
				$G_{tx}$ & Transmitting gain & 0 dBi \\ \hline
				$G_{rx}$ & Reception gain & 0 dBi \\ \hline
				$P_n$ & Noise level & -95 dBm\\ \hline
				CCA & Clear channel assessment & -62 dBm\\ 
				\hline
				\hline
				\textbf{Action number} & \textbf{Transmission power} & \textbf{Channel number} \\ \hline
				 1 & 1 dBm & [36,40] \\ \hline
				 2 & 1 dBm & [44,48] \\ \hline
				 3 & 1 dBm & [36,40,44,48] \\ \hline
				 4 & 20 dBm & [36,40] \\ \hline
				 5 & 20 dBm & [44,48] \\ \hline
				 6 & 20 dBm & [36,40,44,48] \\ \hline
		\end{tabular}}
	}
	\label{tbl:simulation_parameters}
\end{table}

Regarding the reward function, we define a common goal for all the WLANs, which refers to maximize the minimum throughput. To allow a collaborative behavior, the resulting throughput of each WLAN, which is obtained by means of the Shannon capacity, is passed to the ML server. However, note that even if the rewards are known, actions are selected independently for each WLAN, as no other information regarding the configurations of the neighboring WLANs is informed. The Shannon capacity expression is shown in \ref{eq:shannon}:

\begin{algorithm}[t]
	\SetKwInOut{Input}{Input}	
	\Input{$\mathcal{A}$: set of possible actions in \{$1, ..., K$\}}
	initialize: $t=0$,  for each arm $k \in \mathcal{A}$, set $\hat{r}_{k} = 0$ and $n_k = 0$ \\
	\While{active}
	{
    	For each arm $k \in \mathcal{A}$, sample $\theta_k(t)$ from normal distribution $\mathcal{N}(\hat{r}_{k}, \frac{1}{n_k + 1})$ \\
    	Play arm $k = \underset{1,...,K}{\text{argmax }} \theta_k(t) $ \\
    	Observe the reward $r_{k,t}$ \\
    	$ \hat{r}_{k,t} \leftarrow \frac{\hat{r}_{k,t}  n_{k,t} + r_{k,t}}{n_{k,t} + 2}$\\
    	$n_{k,t} \leftarrow n_{k,t} + 1$\\
    	$t \leftarrow t + 1$
	}
	\caption{Implementation of Thompson Sampling for WLANs}
	\label{alg:thompsons}
\end{algorithm}

\begin{figure*}[b]
     \begin{subfigure}[]{0.3\textwidth}
         \centering
         \includegraphics[ width=.8\textwidth]{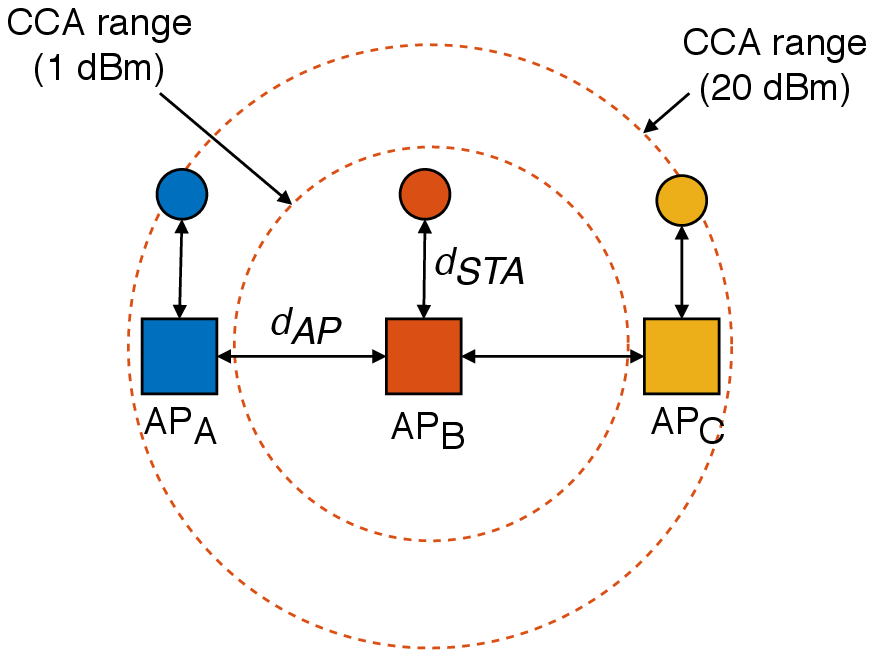}
         \caption{}
         \label{fig:scenario3}
     \end{subfigure}
	 \begin{subfigure}[]{0.3\textwidth}
         \centering
         \includegraphics[ width=\textwidth]{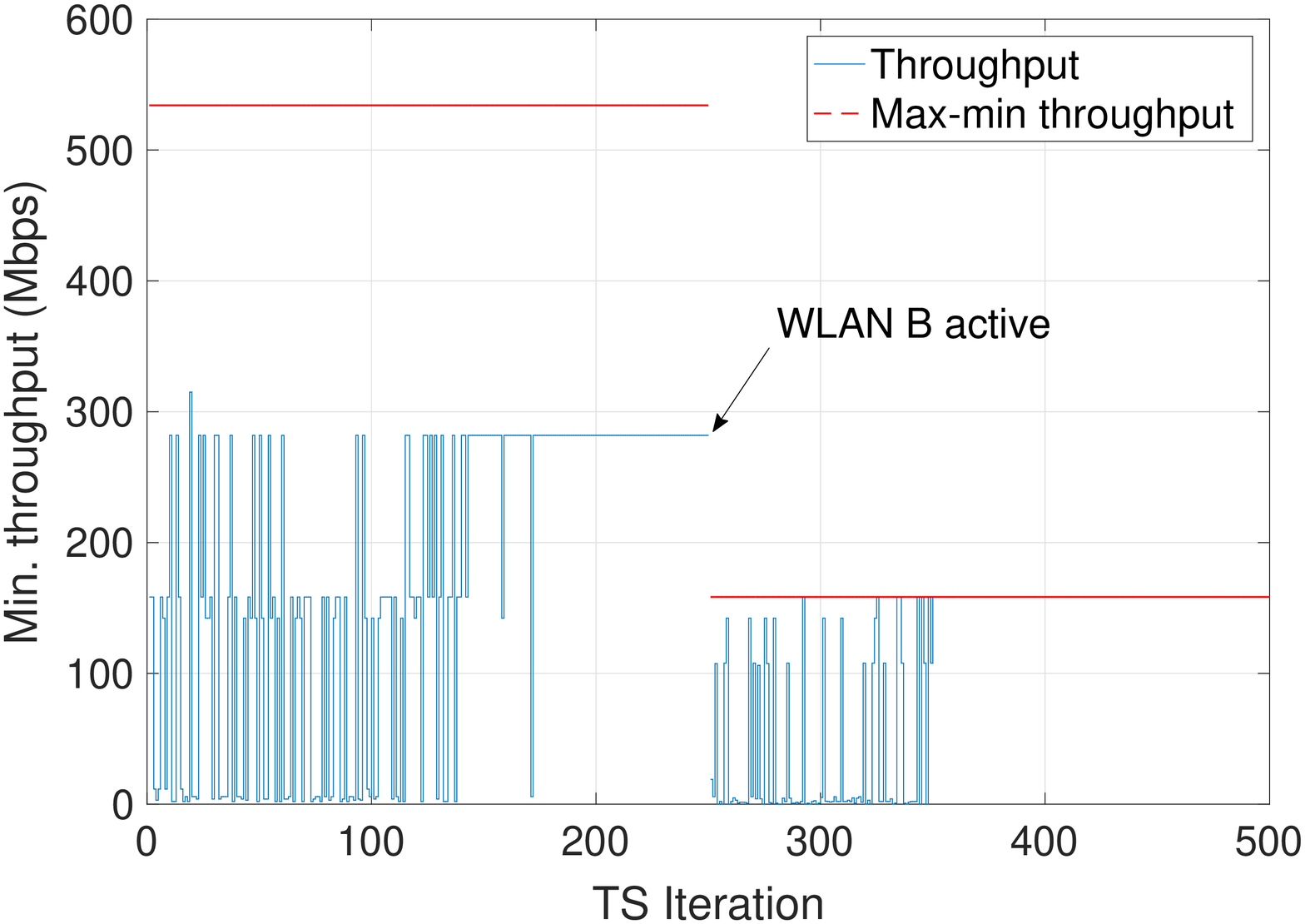}
         \caption{}
         \label{fig:Evolution3}
     \end{subfigure}
     \begin{subfigure}[]{0.3\textwidth}
         \centering
         \includegraphics[ width=\textwidth]{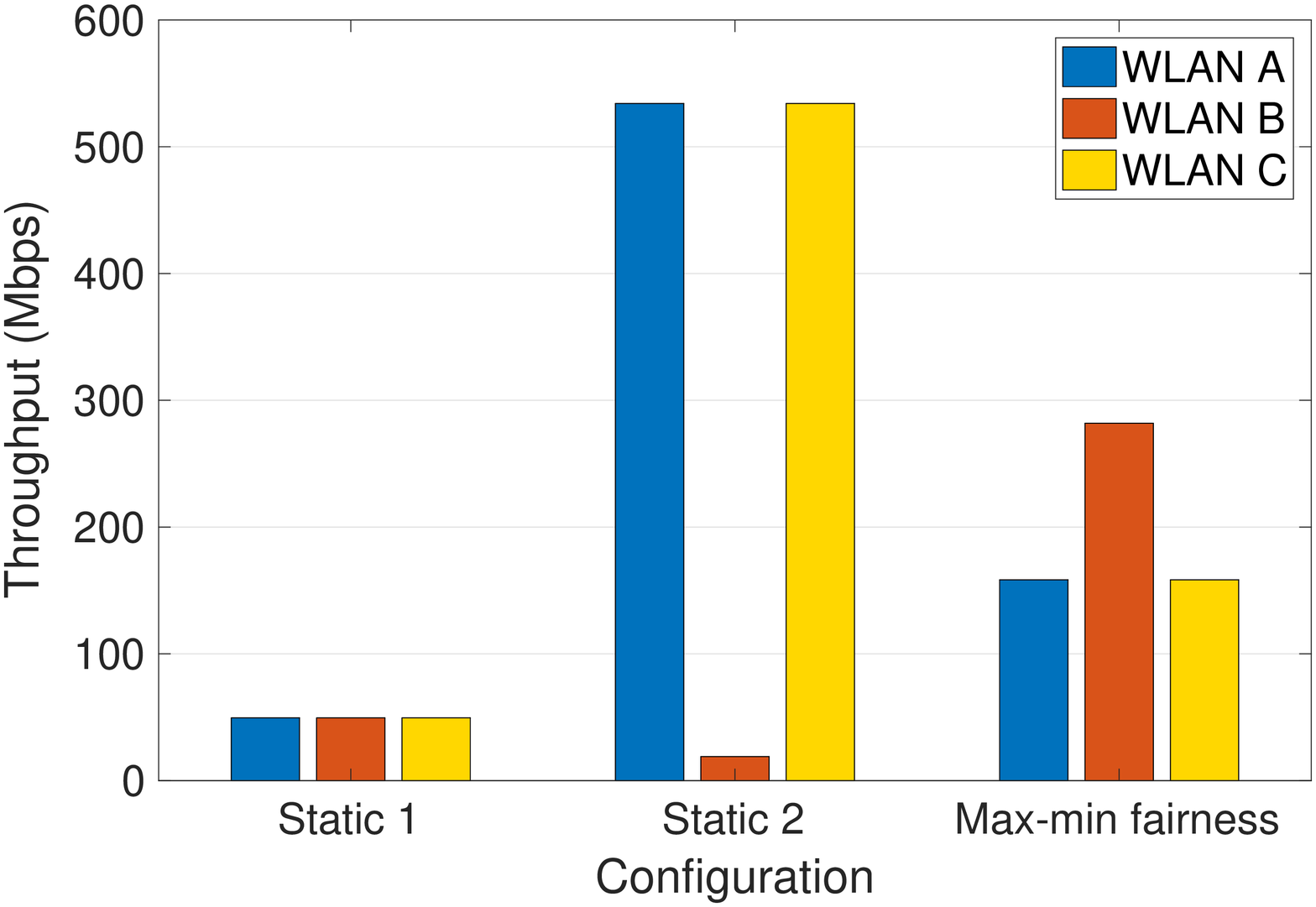}
         \caption{}
         \label{fig:Hist3}
     \end{subfigure}
	\caption[ ]
	{\small Use case with 3 WLAN. (a) Scenario considered. (b) Minimum throughput evolution. (c) Throughput per WLAN corresponding to different action settings. Static 1: All WLAN select action 1 (conservative). Static 2: All WLAN select action 6 (aggressive). Optimal configuration: WLANs A\&C select action 1, whereas WLAN B selects action 5.}
	\label{fig:experiment2}
\end{figure*}

\begin{equation}
\centering
C = B\cdot \log_{2}(1+\text{SINR})
\label{eq:shannon}
\end{equation}
where $B$ is the channel bandwidth, and the SINR is the signal-to-interference-plus-noise ratio given by SINR = $\frac{P_{s}}{P_{n}+P_{i}}$. Here, the $P_{n}$ and $P_{i}$ refer to the noise and interference levels respectively, whereas the $P_{s}$ refers to the signal level received at the AP, which is calculated through the path loss model proposed in \cite{medbo2000simple} that is given in \ref{eq:PL}. This path loss model is simple but accurate, and it is used for 5GHz systems in indoor environments:

\begin{equation}
\centering
L_{prop}(d) = \text{FSL} + \alpha \cdot d
\label{eq:PL}
\end{equation}
where FSL are the well-known free space losses at distance $d$, and $\alpha = 0.44$ dB/m is the constant attenuation per unit of path length. The different simulation parameters taken into account are described in table \ref{tbl:simulation_parameters}.

\begin{figure*}[t]
     \begin{subfigure}[]{0.3\textwidth}
         \centering
         \includegraphics[ width=.7\textwidth]{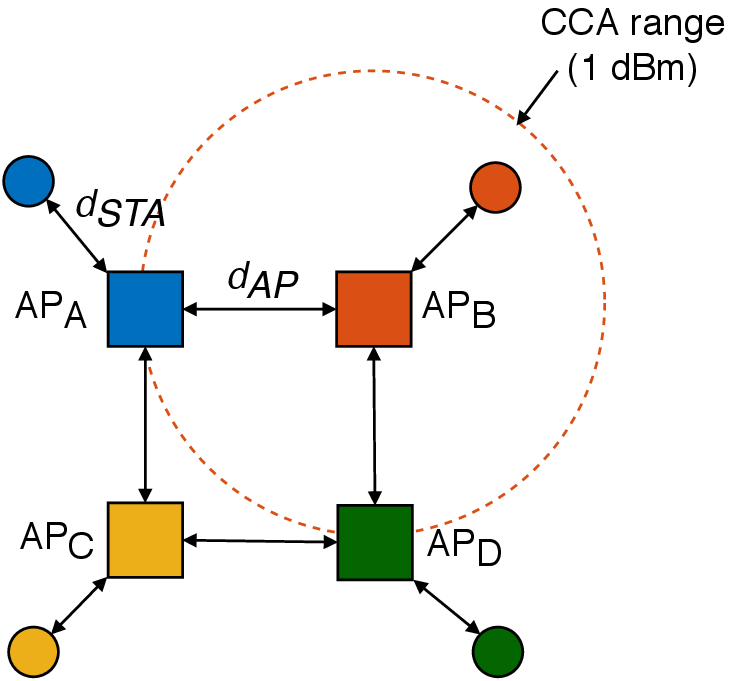}
         \caption{}
         \label{fig:scenario4}
     \end{subfigure}
	 \begin{subfigure}[]{0.3\textwidth}
         \centering
         \includegraphics[ width=\textwidth]{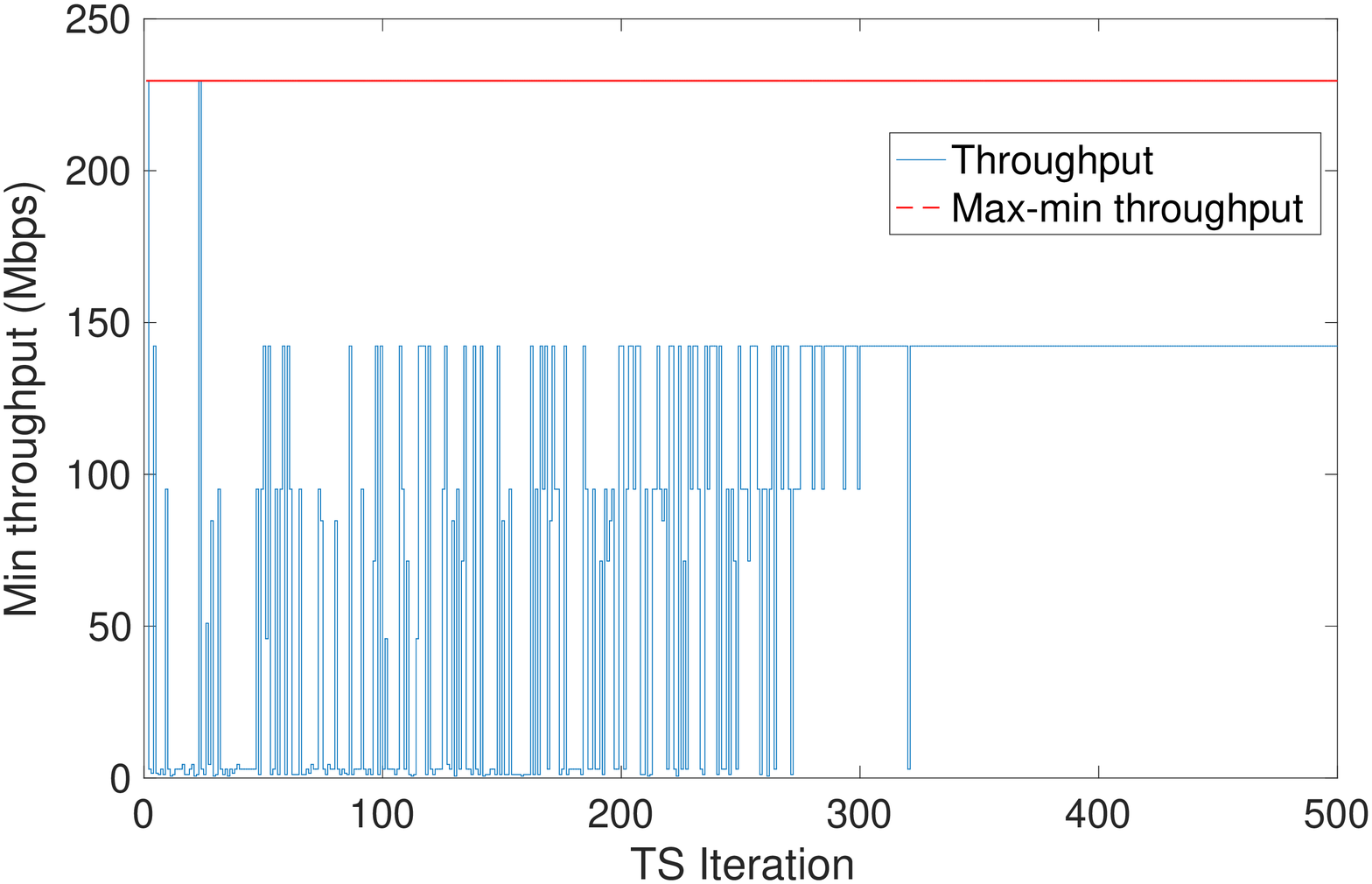}
         \caption{}
         \label{fig:minTh4wlan}
     \end{subfigure}
     \begin{subfigure}[]{0.3\textwidth}
         \centering
         \includegraphics[ width=\textwidth]{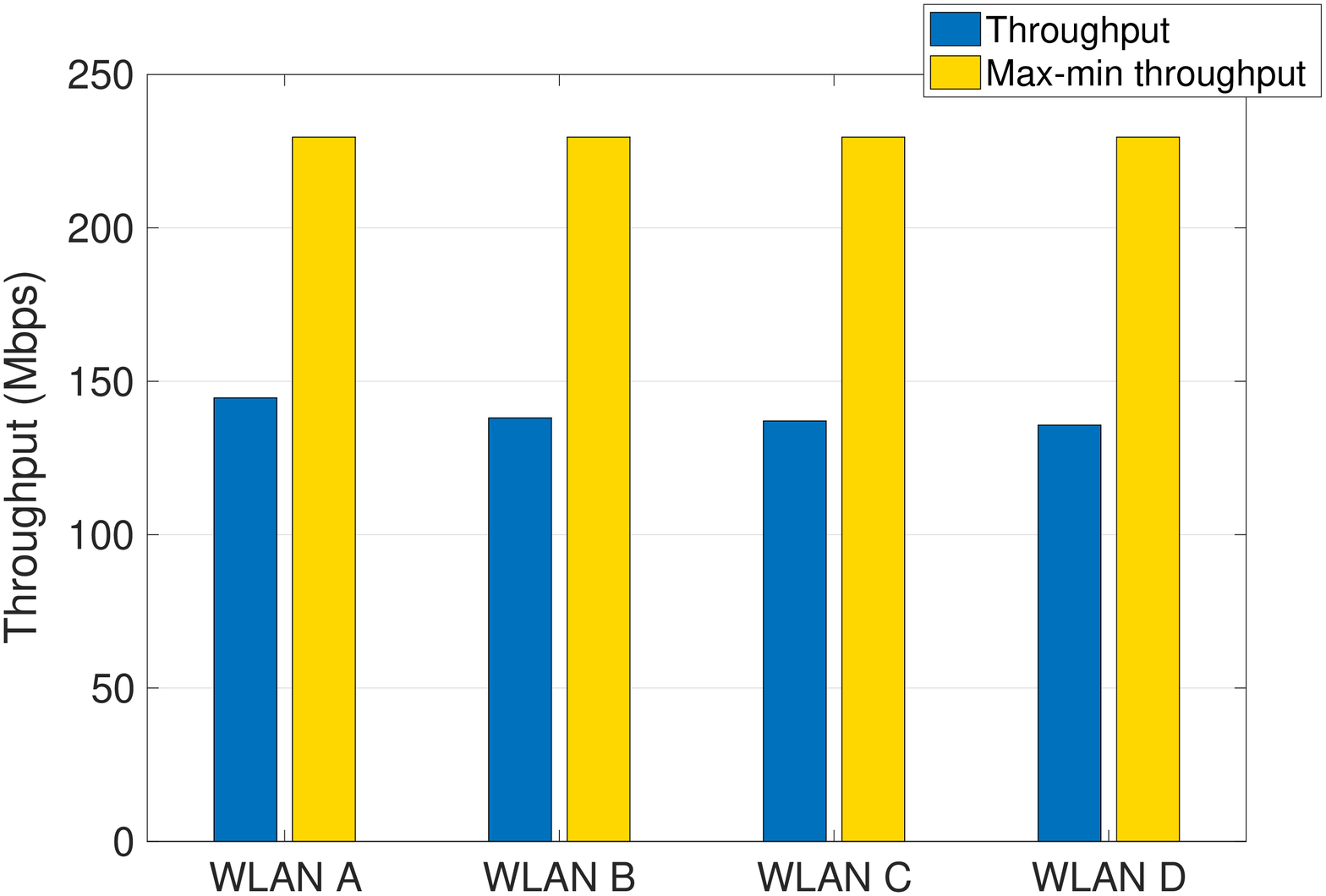}
         \caption{}
         \label{fig:HtH4wlan}
     \end{subfigure}
	\caption[ ]
	{\small Use case with 4 WLANs. (a) Scenario considered. (b) Minimum throughput evolution. (c) Throughput histogram.}
	\label{fig:experiment3}
\end{figure*}
\subsection{Full overlapping WLANs}

In this first scenario, which is presented in Figure \ref{fig:scenario2}, we consider 2 WLANs that fully overlap. The parameters $d_{STA}$ and $d_{AP}$ are set to 5 m. Either choosing 1 dBm or 20 dBm, both APs will be inside the CCA range of its neighbor. At the end of the simulation, we observe that both WLANs reached the optimal configuration. The two WLANs selected the maximum transmission power, and a different channel scheme as it can be seen in Figure \ref{fig:actions2}. Besides, we can see that actions containing the whole set of channels have been explored but discarded, as they were only beneficial for one WLAN in detriment of the other. Moreover, in Figure \ref{fig:Evolution2}, we can observe that the max-min throughput converges into a collaborative solution before iteration $200$, discarding selfish decisions. Regarding the transmission power, both networks decide to use the maximum allowable as using the lower value does not reduce the contention between the two WLANs.

\subsection{Partial overlapping WLANs}

In this scenario, we want to tackle a non-stationary scenario by simulating changing conditions. For this purpose, we deploy three partially overlapping WLANS (Figure \ref{fig:scenario3}),  which activation time is different. Then, WLAN A and C are activated since the beginning, whereas WLAN B is activated at iteration $250$. The parameters $d_{STA}$ and $d_{AP}$ are set to 5 m. Figure \ref{fig:Evolution3} shows the obtained results, and how TS fails at reaching the best possible configuration for the first $250$ iterations, so WLANs A and C end up choosing different channel ranges in order to avoid interference. In this particular case, the optimal configuration is not found since it requires both WLANs to choose the optimal action simultaneously (i.e., minimum transmit power and the entire channel range). Moreover, in case that only one of the WLANs chooses the optimal one, it becomes vulnerable if the other WLAN uses maximum transmit power, thus leading to a low collaborative reward. On the other hand, when WLAN B becomes active, the three WLANs are able to choose the optimal configuration. Note that at iteration $250$, the previous knowledge is discarded since the network state has changed. In Figure \ref{fig:Hist3}, we have performed a comparison between applying learning, and leaving WLANs with an static configuration. We show that this kind of techniques can minimise the appearance of problems such as the flow starvation. For instance, from the scenario presented in Figure \ref{fig:scenario3}, we can see that WLAN B will suffer flow starvation as the other WLANs will get most of the time to transmit. If we do not apply any mechanism and we explore the different available actions, we find that applying a conservative action (i.e., minimum power and minimum bandwidth) will lead to downgrade the performance of the three WLANs but maintaining the fairness. On the contrary, if we apply and aggressive solution, WLAN B barely transmits. As a result, none of the previous solutions solves the situation without diminishing the performance of several WLANs, nor making the network unfair.

\subsection{Grid scenario}

Lastly, we have studied the behaviour of the proposed solution in a grid scenario, which is depicted in Figure \ref{fig:scenario4}. The parameters $d_{STA}$ and $d_{AP}$ are set as $\sqrt8$ m and 5 m respectively. Here, we intend to see the interactions between multiple neighbors, and how the decisions of others affect the action-selection process. For this scenario, finding the optimal configuration in a decentralized way is unlikely to occur, since it requires that all WLANs choose the optimal action simultaneously. Therefore, there is a narrow window of possibilities for that to happen. In case that only one of the WLANs chooses the optimum, it becomes vulnerable and so leading to a low collaborative reward. However, as shown in Figure \ref{fig:minTh4wlan}, the learning algorithms reach a solution that is fair. Note that as the number of nodes increases, the convergence time increases too. So, the more nodes we have, the later we converge into a solution when considering a collaborative reward. Figure \ref{fig:HtH4wlan} shows a comparison among optimal and achieved throughput per WLAN.

\section{Conclusions} \label{sec:conclusions}

In this article, we have shown that new networking paradigms, such as the presented SDN and SDWN, are grabbing attention from academia and research institutions, with a clear aim to be used in next generation of WLAN deployments. Besides, big data mining and machine learning techniques are also raising attention due to their ability to use collected information for improving network management. In this regard, we have performed different study cases to analyse the behavior of ML over wireless networks for management purposes. ML and SDWN can be perfectly combined to achieve better performance, as the results obtained prove that there is a clear improvement over the pre-defined configurations. However, further research must be carried out in order to quantify the different drawbacks and trade-offs that exist, such as the negative effects that greater network delays can have in the overall network performance.

\section*{Acknowledgment}

This  work  has  been  partially  supported  by  the  Spanish Ministry of Economy and Competitiveness under the Maria de Maeztu  Units  of  Excellence  Programme  (MDM-2015-0502), by the Catalan Government under grant 2017-SGR-1188, by the Spanish Government under grant PGC2018-099959-B-I00 (MCIU/AEI/FEDER,UE), and  by  a  Gift from the Cisco University Research Program (CG\#890107, Towards Deterministic Channel Access in High-Density WLANs) Fund, a corporate advised fund of Silicon Valley Community Foundation. The work done by S. Barrachina-Mu\~noz is supported by a FI grant from the Generalitat de Catalunya.


\bibliographystyle{unsrt}
\bibliography{bib}

\begin{thebibliography}{10}

\bibitem{index2015cisco}
Cisco Visual~Networking Index.
\newblock Cisco visual networking index: Global mobile data traffic forecast
  update, 2016-2021.
\newblock {\em White Paper}, March, 2017.

\bibitem{bellalta2016ieee}
B.~Bellalta.
\newblock {IEEE} 802.11ax: High-efficiency {WLANs}.
\newblock {\em IEEE Wireless Communications}, 23(1):38--46, 2016.

\bibitem{yap2010openroads}
Kok-Kiong Yap, Masayoshi Kobayashi, Rob Sherwood, Te-Yuan Huang, Michael Chan,
  Nikhil Handigol, and Nick McKeown.
\newblock Openroads: Empowering research in mobile networks.
\newblock {\em ACM SIGCOMM Computer Communication Review}, 40(1):125--126,
  2010.

\bibitem{suresh2012towards}
Lalith Suresh, Julius Schulz-Zander, Ruben Merz, Anja Feldmann, and Teresa
  Vazao.
\newblock Towards programmable enterprise wlans with odin.
\newblock In {\em Proceedings of the first workshop on Hot topics in software
  defined networks}, pages 115--120. ACM, 2012.

\bibitem{schulz2015opensdwn}
Julius Schulz-Zander, Carlos Mayer, Bogdan Ciobotaru, Stefan Schmid, and Anja
  Feldmann.
\newblock Opensdwn: programmatic control over home and enterprise wifi.
\newblock In {\em Proceedings of the 1st ACM SIGCOMM Symposium on Software
  Defined Networking Research}, page~16. ACM, 2015.

\bibitem{yiakoumis2015behop}
Yiannis Yiakoumis, Manu Bansal, Adam Covington, Johan van Reijendam, Sachin
  Katti, and Nick McKeown.
\newblock Behop: A testbed for dense wifi networks.
\newblock {\em ACM SIGMOBILE Mobile Computing and Communications Review},
  18(3):71--80, 2015.

\bibitem{moura2015ethanol}
Henrique Moura, Gabriel~VC Bessa, Marcos~AM Vieira, and Daniel~F Macedo.
\newblock Ethanol: Software defined networking for 802.11 wireless networks.
\newblock In {\em Integrated Network Management (IM), 2015 IFIP/IEEE
  International Symposium on}, pages 388--396. IEEE, 2015.

\bibitem{riggio2015programming}
Roberto Riggio, Mahesh~K Marina, Julius Schulz-Zander, Slawomir Kuklinski, and
  Tinku Rasheed.
\newblock Programming abstractions for software-defined wireless networks.
\newblock {\em IEEE Transactions on Network and Service Management},
  12(2):146--162, 2015.

\bibitem{vestin2013cloudmac}
Jonathan Vestin, Peter Dely, Andreas Kassler, Nico Bayer, Hans Einsiedler, and
  Christoph Peylo.
\newblock Cloudmac: towards software defined wlans.
\newblock {\em ACM SIGMOBILE Mobile Computing and Communications Review},
  16(4):42--45, 2013.

\bibitem{schulz2014aeroflux}
Julius Schulz-Zander, Nadi Sarrar, and Stefan Schmid.
\newblock Aeroflux: A near-sighted controller architecture for software-defined
  wireless networks.
\newblock In {\em Presented as part of the Open Networking Summit $\{$ONS$\}$},
  2014.

\bibitem{yan2015aetherflow}
Muxi Yan, Jasson Casey, Prithviraj Shome, Alex Sprintson, and Andrew Sutton.
\newblock Aetherflow: principled wireless support in sdn.
\newblock In {\em 2015 IEEE 23rd International Conference on Network Protocols
  (ICNP)}, pages 432--437. IEEE, 2015.

\bibitem{patro2015coap}
Ashish Patro and Suman Banerjee.
\newblock Coap: A software-defined approach for home wlan management through an
  open api.
\newblock {\em ACM SIGMOBILE Mobile Computing and Communications Review},
  18(3):32--40, 2015.

\bibitem{clark2003knowledge}
David~D Clark, Craig Partridge, J~Christopher Ramming, and John~T Wroclawski.
\newblock A knowledge plane for the internet.
\newblock In {\em Proceedings of the 2003 conference on Applications,
  technologies, architectures, and protocols for computer communications},
  pages 3--10. ACM, 2003.

\bibitem{mestres2017knowledge}
Albert Mestres, Alberto Rodriguez-Natal, Josep Carner, Pere Barlet-Ros, Eduard
  Alarc{\'o}n, Marc Sol{\'e}, Victor Munt{\'e}s-Mulero, David Meyer, Sharon
  Barkai, Mike~J Hibbett, et~al.
\newblock Knowledge-defined networking.
\newblock {\em ACM SIGCOMM Computer Communication Review}, 47(3):2--10, 2017.

\bibitem{azzouni2017neutm}
Abdelhadi Azzouni and Guy Pujolle.
\newblock Neutm: A neural network-based framework for traffic matrix prediction
  in sdn.
\newblock {\em arXiv preprint arXiv:1710.06799}, 2017.

\bibitem{kim2016congestion}
Seonhyeok Kim, Jaehyeok Son, Ashis Talukder, and Choong~Seon Hong.
\newblock Congestion prevention mechanism based on q-leaning for efficient
  routing in sdn.
\newblock In {\em Information Networking (ICOIN), 2016 International Conference
  on}, pages 124--128. IEEE, 2016.

\bibitem{tang2016deep}
Tuan~A Tang, Lotfi Mhamdi, Des McLernon, Syed Ali~Raza Zaidi, and Mounir
  Ghogho.
\newblock Deep learning approach for network intrusion detection in software
  defined networking.
\newblock In {\em Wireless Networks and Mobile Communications (WINCOM), 2016
  International Conference on}, pages 258--263. IEEE, 2016.

\bibitem{wilhelmi2019collaborative}
Francesc Wilhelmi, Cristina Cano, Gergely Neu, Boris Bellalta, Anders Jonsson,
  and Sergio Barrachina-Mu{\~n}oz.
\newblock Collaborative spatial reuse in wireless networks via selfish
  multi-armed bandits.
\newblock {\em Ad Hoc Networks}, 88:129--141, 2019.

\bibitem{barrachina2019dynamic}
Sergio Barrachina-Mu\~noz, Francesc Wilhelmi, and Boris Bellalta.
\newblock {Dynamic channel bonding in spatially distributed high-density
  WLANs}.
\newblock {\em (In press) IEEE Transactions on Mobile Computing}, 2019.

\bibitem{medbo2000simple}
Jonas Medbo and J-E Berg.
\newblock Simple and accurate path loss modeling at 5 ghz in indoor
  environments with corridors.
\newblock In {\em Vehicular Technology Conference, 2000. IEEE-VTS Fall VTC
  2000. 52nd}, volume~1, pages 30--36. IEEE, 2000.

\end{thebibliography}

\end{document}